\documentclass[12pt,preprint]{aastex}
\shorttitle{4.8- and 8.6-GHz Images of the LMC}
\shortauthors{Dickel et al.}

\begin{document}

\title{A 4.8- and 8.6-GHz Survey of the Large Magellanic Cloud: I The Images}

\author{John. R.  Dickel}
\affil{Astronomy Department, University of Illinois at Urbana-Champaign, 
1002 West Green Street, Urbana IL 61801, USA}

\author{Vincent J. McIntyre}
\affil{CSIRO Australia Telescope National Facility, Box 76, Epping NSW 1710, 
Australia} 

\author{Robert A. Gruendl}
\affil{Astronomy Department, University of Illinois at Urbana-Champaign, 
1002 West Green Street, Urbana IL 61801, USA}

\and

\author{Douglas K. Milne}
\affil{CSIRO Australia Telescope National Facility, Box 76, Epping NSW 1710, 
Australia}

\begin{abstract}

Detailed 4.8- and 8.6-GHz radio images of the entire Large Magellanic
Cloud with half-power beamwidths of 33$\arcsec$ at 4.8 GHz and 20$\arcsec$ at
8.6 GHz have been obtained using the Australia Telescope Compact Array.
A total of 7085 mosaic positions were used to cover an area of 6$^{\circ}$
on a side.  Full polarimetric observations were made.  These images
have sufficient spatial resolution ($\sim$8 and 5 pc, respectively) and
sensitivity (3 $\sigma$ of 1 mJy beam$^{-1}$) to identify most of the 
individual
SNRs and H II regions and also, in combination with available data from
the Parkes 64-m telescope, the structure of the smooth emission in that
galaxy.  In addition, limited data using the sixth antenna at 4.5 to 6-km
baselines are available to distinguish bright point sources ($<$ 3 and
2 arcsec, respectively) and to help estimate sizes of individual
sources smaller than the resolution of the full survey.  The resultant
database will be valuable for statistical studies and comparisons
with x-ray, optical and infrared surveys of the LMC with similar
resolution.

\end{abstract}


\section{INTRODUCTION}

The Large Magellanic Cloud (LMC) is an excellent laboratory for
investigation of the processes of star birth and death, and particularly
the interactions of the stellar population and various components of
the interstellar medium (ISM).  To measure many properties of the sources, 
such as luminosity and physical 
size, we must know their distances. This is  hard to do for objects in 
the Milky Way because most of the measured quantites are 
themselves dependent on the distance and there is also obscuring material 
within the Milky Way itself which can blank our view and distort optical 
measurements.  In the LMC  all sources of emission are at essentially the
same known distance of 50~kpc.

Radio continuum observations can be used to measure both thermal emission 
from H~II regions and polarized synchrotron emission from supernova remnants 
(SNRs).  The thermal emission has a nearly flat radio spectrum whereas the 
synchrotron emission, from relativistic electrons accelerated in shocks and 
the magnetic fields of SNRs, has a power-law spectrum with brighter emission 
at lower frequencies. 
Three important sources of stellar energy in the ISM are UV radiation,
fast stellar winds, and supernovae.  Thus the radio observations   
supply a vital component in studies of these
energy sources, by tracing the contributions from the photoionized and
shock-ionized material and (through polarimetry) showing the orientation of 
magnetic field lines, even deep within highly obscured regions like molecular
clouds.  In combination with optical line emission \citep{s99a} and X-ray 
images \citep{s99b} from the
new generation surveys at those wavelengths, high-resolution
radio continuum imaging will help to construct a complete physical picture 
of stellar feedback in the Magellanic Clouds.

We have completed 4.8- and 8.6-GHz observations of the LMC to produce 
images $6^{\circ}$ on a side, including all 
the bright features in that galaxy.  They have resolutions of 33\arcsec ~at 
4.8~GHz (20\arcsec ~at 8.6~GHz) or about 8 pc
(5~pc) at the 50~kpc distance of the LMC.  With these moderate-resolution 
data in hand, one can choose smaller areas for followup imaging, and reuse 
these relatively short-spacing data.
 
The noise level of the system provides 3-$\sigma$ measurements at 
1~mJy beam$^{-1}$ and if the source fills the 33$\arcsec$ beam at 4.8 GHz, 
this value corresponds 
to a surface brightness at of 4$\times$10$^{-22}$ W m$^{-2}$ Hz$^{-1}$ 
sr$^{-1}$ or for a mean SNR spectral index of $-0.5$,
a 1 GHz surface brightness of 1$\times$10$^{-21}$ W m$^{-2}$ 
Hz$^{-1}$sr$^{-1}$.
While this limit is ten times the brightness of the faintest SNR in
the Milky Way (G182.4+4.3, \cite*{k98}) it would include 90\%
of all the Milky Way SNRs cataloged by Green
(http://www.mrao.cam.ac.uk/surveys/snrs/).

\section{OBSERVATIONS}

\subsection{Telescope Configurations}

The observations were made simultaneously at 4.80 and 8.64 GHz using the 
Australia Telescope Compact Array (ATCA) consisting of five moveable 22-meter 
antennas and a sixth one fixed at a distance of 3 km from the end of the 3-km 
track available for the moveable antennas \citep{at92}.  For 
this program, two configurations of the inner five antennas were used -- 367 
and 352 meters.  Together they provided nineteen antenna spacings between 31 
and 367 meters separated by 15.3 meters with four gaps of 30.6 meters.  This 
setup gave nearly complete coverage of the available spacings and thus a well 
sampled beam.  The surface brightness of the dominant H~II region, 
molecular cloud complex 30~Doradus is over 1.3 Jy beam$^{-1}$ at 4.8 GHz, the 
SNR N132D is over 0.4 Jy beam$^{-1}$ and several other sources approach that 
value.  The desire to detect sources as faint as 1 mJY beam$^{-1}$ thus
necessitated a dynamic range 
of over 1000/1.  Extending the spacings of the antennas to improve the 
resolution would have resulted in sparser sampling of the spatial frequency 
plane ($uv$-plane) with increased sidelobe levels and a lower dynamic range.   
The declination of the LMC is close enough to 
the south pole that the beam for this east-west array is nearly circular. 

The complex southeastern quarter of the galaxy was observed with an 
additional 375-meter configuration.  This added three additional spacings to 
improve the beam pattern somewhat and also increased the sensitivity in that 
area by $\sqrt{3/2}$.

Data from the sixth antenna were also recorded to give a set of longer 
spacings reaching resolutions of 3\arcsec ~at 4.8 GHz and 2\arcsec ~at 
8.6 GHz.
Without intermediate spacings, it is not possible to construct a proper beam 
at this resolution but correlation of the data from antenna 6 with the others 
can be used to compare the flux density of a given source with that determined 
from the more complete synthesis with lower resolution to see what fraction of
the source is unresolved on the complete image at the lower resolution.  This
procedure helps in the identification of background sources seen through 
the LMC.  Full images with the higher resolution data have not been 
constructed but readers may obtain the data for any given position and make 
the images themselves. 

The nominal bandwidth of the system at each frequency is 128 MHz split into 
thirty-two 4 MHz channels.  No narrow band interference was detected in the 
individual channels.  To 
account for edge effects in the filters, the inner 104 MHz was used at each 
band.

\subsection{Observing Procedure}

The observations were carried out during five extended runs with the ATCA.  
The 375-configuration was observed in March 2001;  the 367-configuration in 
March 2002 and December 2002;  and the 352-configuration in December 2001 
and March 2003.  The total observing time for this project was 496 hours. 

The final image encompasses an area 6$^{\circ}$ on a side centered at 
$05^h19^m00^s$ and $-69^{\circ}00\arcmin00\arcsec$ (J2000).  
It includes almost all of the important objects in the LMC.  To cover an 
area that large, a mosaic of images at 7085 
individual pointing positions was made.  This number included a few makeups 
for mis-pointings, increased coverage particularly toward 30~Dor, etc.  
The pointings were arranged in a so-called
 ``hexagonal pattern'' of equilateral triangles such that each point 
was equidistant from six surrounding ones.  With circular symmetry in the 
beam this arrangement allows a slightly larger spacing between points than the 
normal 1/2 HPBW for a standard rectangular pattern.  The adopted separation 
between points was $5\farcm 0$. This distance was a compromise that somewhat 
oversampled at 4.8 GHz (HPBW $9\farcm 8$) and somewhat undersampled at 8.6 GHz 
($5\farcm 4$).  

\begin{figure*}
\epsscale{1.0}
\plotone{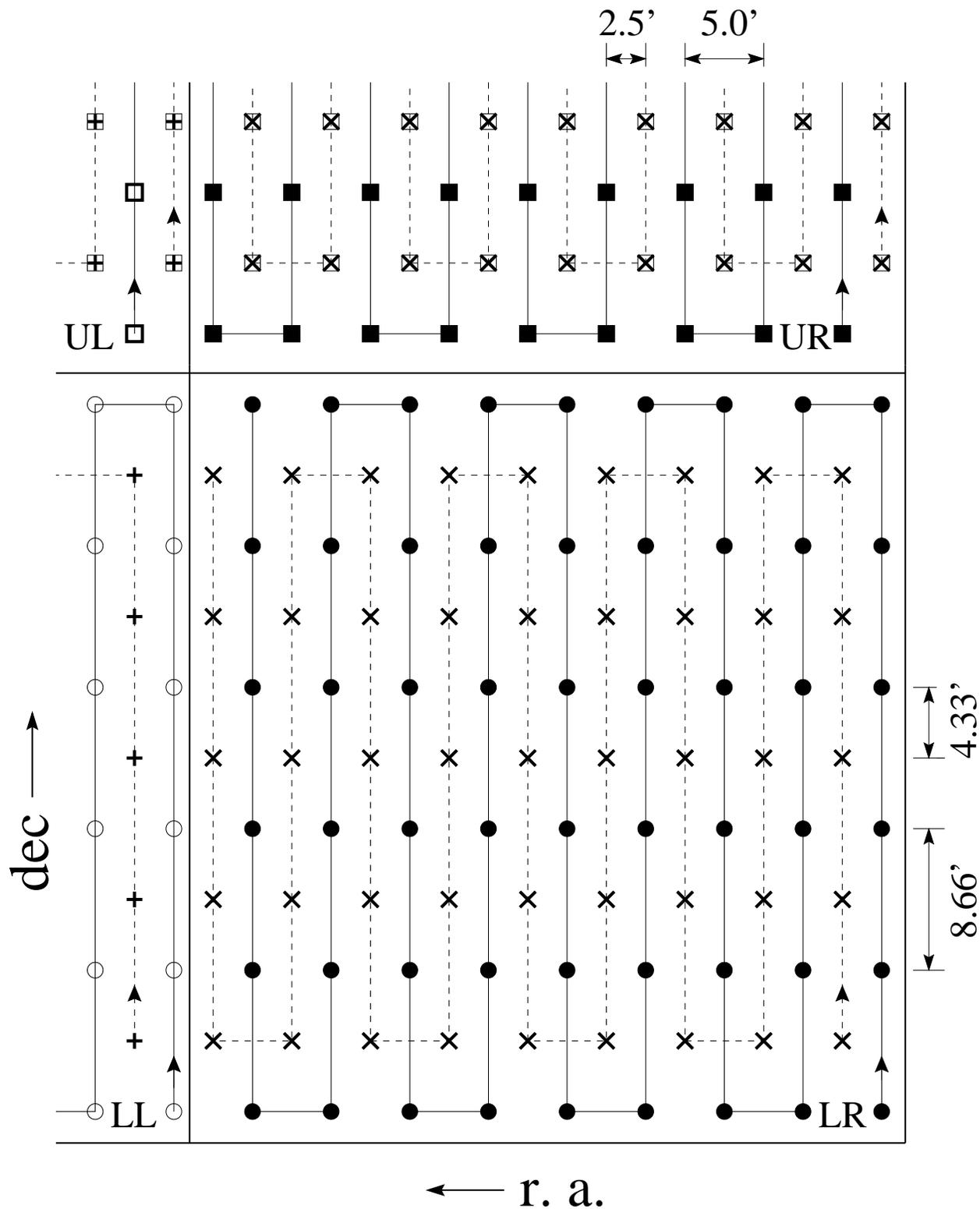}
\caption{Sample observing patterns for the lower right field within one tile 
and parts of the adjoining fields.  The circles and boxes represent the 6-row 
patterns and the Xs and crosses represent the 5-row patterns within the field. 
The separations between points are shown.}
\label{fig-pattern}
\end{figure*}

The entire area was broken into 16 tiles, each $1.5^{\circ}$ on a side.  Each 
tile was observed during a full 12-hour observing ``day'' in each 
configuration.  There were four fields within one tile and two observing 
patterns within a field to minimize move time and provide as much hour angle 
coverage per pointing as practical.  The observing patterns were laid out as 
shown in Figure~\ref{fig-pattern} with 9 colums of 6 vertical points 
alternating with 9 columns of 5 
points each.  Each point was observed for two 10 -- 12 second integrations 
(depending on whether the observing ``day'' was 12 or 13 hours long plus 
compensation for any lost time) before moving to the next point in the pattern.
The eight patterns for a given tile were staggered throughout an $\sim$ 3-hour 
period so that adjacent columns were observed at hour angles separated by 
$\sim$  1 1/2 hours.  The full 3-hour sequence of patterns, including 
calibration 
observations, was cycled four times to give eight reasonably uniformly spaced 
hour angles for each location in the galaxy.  The total integration time per 
pointing was thus $\ge 8 \times 20^s$.

\begin{figure*}
\epsscale{1.0}
\plotone{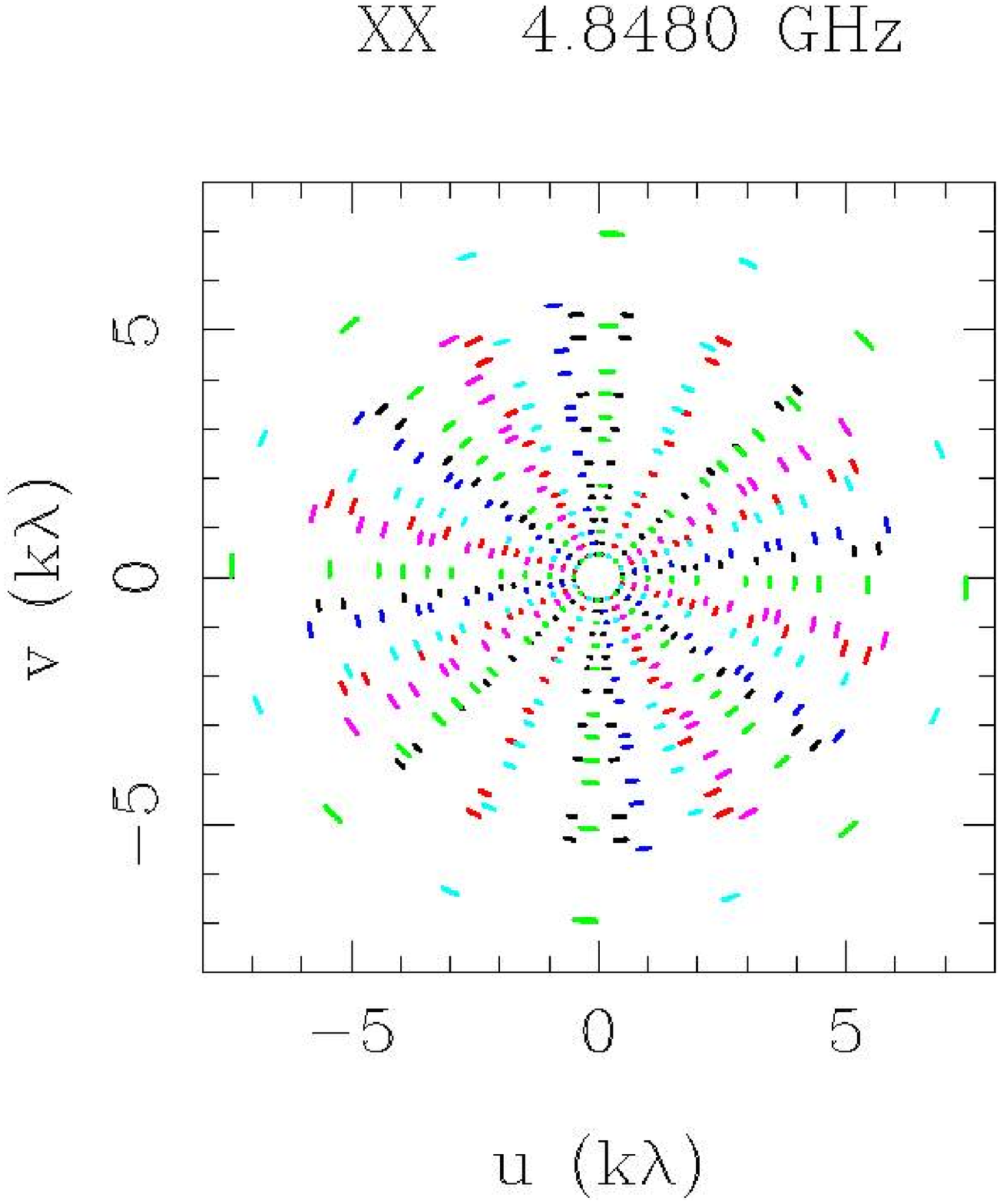}
\caption{Coverage of the spatial frequency ($uv$) plane at 4.8 GHz for both 
patterns covering one field with all arrays.  Note that this covers only the 
inner 400 meters and does not include the data fron the distant 6-km antenna.}
\label{fig-coverage}
\end{figure*}

The observations with the second 
telescope configuration were obtained at hour angles that interleaved
with those from the first configuration, double 
the final number of hour angles.  Figure~\ref{fig-coverage} is a plot of the 
$uv$ coverage of the 
field including the bright H~II region complex 30~Dor \citep{b01} which
included all three array configurations.  Each short arc is the coverage for
one baseline for the duration of one observing pattern.    

The individual fields and tiles were positioned such that the pointing 
separations were maintained over the tile boundaries.  Because the
right ascension
separations are measured in hours of right ascension rather than
angular distance on  
the sky, the right ascension separation varies with declination.  In practice, 
the separation was held constant in right ascension for each field and set 
to be correct for the 
northern most points in the field so that the separations were always the 
designated 5$\arcmin$ or slightly less.

Absolute calibration was obtained by observations of PKS B1934-638 for which 
flux densities of 5.8 and 2.8 Jy at 4.8 and 8.6 GHz, respectively, were 
adopted from the memo by Reynolds 
(http://www.atnf.csiro.au/observers/memos/d96783\~{}1.pdf).
PKS B0454-810 was used for phase and pointing calibration.

\section{DATA REDUCTION}

\subsection{Total Intensity}

The data were split into the two frequency bands, edited and calibrated using 
standard MIRIAD \citep*{stw95} routines.  To compromise between the desire for 
full sensitivity from natural weighting 
of the data and high resolution with a good beam pattern from uniform 
weighting, a Briggs' (\citeyear{b95}) robust weighting of 0 was adopted. The 
resultant half-power beamwidths were 33\arcsec ~at 4.8 GHz and 
20\arcsec ~at 8.6 GHz.

Because the baseline lengths are all multiples of 15~meters, there is a bright 
grating ring with a radius of 14\farcm3 at 4.8  GHz (and 8\farcm0 at 
8.6 GHz) around all strong sources.  To remove this ring by the CLEAN process 
it is necessary to construct images large enough to encompass this ring which 
lies well outside the primary beam of the individual antennas.  Thus the 
usual mosaic mode of imaging in MIRIAD could not be used. Instead, each
pointing was imaged, 
CLEANed and restored individually before combining into the final mosaic of 
the entire galaxy.  The individual images were 500 8\arcsec ~pixels on a side 
at 4.80 GHz and 500 5\arcsec ~pixels at 8.6 GHz.

To assess the noise level on the final image, statistics were determined for 
six different randomly-placed boxes with areas of about 900 square
arcmin containing no sources brighter than 5 mJy beam$^{-1}$.
The average rms noise was 0.28 mJy 
beam$^{-1}$ at 4.8 GHz and 0.50 mJy beam$^{-1}$ at 8.6 GHz.  
Because of the 
overlapping mosaic positions, the effective integration time at any location 
is roughly four times the value of $\sim 160$ sec per pointing position.  
Indeed 
the measured noise levels are close to the theoretical values of 0.31 
mJy beam$^{-1}$ and 0.38 mJy beam$^{-1}$ at 4.8 and 8.6 GHz, respectively 
for continuum observations with the chosen arrays and an integration time of 
640 sec (see http://www.atnf.csiro.au/observers/docs/at\_sens/).  In the 
southeastern quadrant where the third array was used, the rms noise levels 
were somewhat better at 0.25 and 0.46 mJy beam$^{-1}$, respectively.  This is 
not quite as much gain as expected purely on the basis of observing time 
because of the significant confusion in this complex part of the LMC.

For the first pass of cleaning the images for each pointing we chose a cutoff 
brightness of 0.5 mJy beam$^{-1}$.  Fewer than 10\%  of the positions did 
not converge at that level; in these cases a higher cutoff was adopted.
All but about fifty 
positions converged at levels less than 0.7 mJy beam$^{-1}$.  The others were 
in complex regions of bright sources and the dynamic range remained greater 
than 500/1.    

The region around 30~Dor contains bright, complex emission which
produces strong sidelobe structures. To improve the dynamic range 
there, we utilized archival ATCA data sets \citep*{l00,l03}
which have near-complete $uv$-plane coverage
for 30~Dor and N~157B, at both frequencies.
We also included some makeup observations in the region of N~160 with
more complete hour angle coverage than the survey observations.

\subsection{Polarization}

Stokes Q, U, and V images were constructed, cleaned, and mosaiced in the 
same manner as the total intensity ones.  Because the square root of the sum 
of the squares of the Q and U stokes intensities will always be positive, 
the resultant polarized intensity images were statistically corrected for the 
Ricean bias to account for noise allowing negative 
intensities \citep*{k86,v65}.  
The Stokes V images, representing circular polarization show no structure 
above the noise level at either frequency.  In general the linearly polarized 
intensities show no cross-polarization leakage above 0.1 mJy beam$^{-1}$ 
except toward the bright thermal source 30~Dor where the polarized intensities 
reach 2 mJy beam$^{-1}$ and 3 mJy beam$^{-1}$ to give leakages of 0.15\% and 
0.2\% at 4.8 and 8.6 GHz, respectively.   

The typical Faraday rotation for objects 
in  the LMC is less than few hundred rad m$^{-2}$ \citep[e.g.,][]{dm95,dm98} 
so the total rotation across the band at 
4.8 GHz would be less than 10$^{\circ}$.  We have thus ignored bandwidth 
smearing in the polarization data.

\subsection{Addition of Parkes Data}

As shown by \citet{Haynes91}, the LMC presents a complex field full of
extended emission. Any interferometric observation aimed at studying
the extended sources in the LMC will thus benefit from a database of complete
short-spacing $uv$ coverage.  The closest spacing between individual ATCA 
antennas of 30 meters cuts off 
those shortest available interferometer spacings.  To include this information,
we have utilized observations from the Parkes 64-meter telescope at 4.75 GHz 
and 8.55 GHz by \citet{Haynes91}.  Both total intensity and polarization 
images were merged at 4.8 GHz but only the total intensity at 8.6 GHz because 
of the lack of sufficient polarized intensity to produce Parkes images at that
frequency.   

The ATCA data were directly observed in J2000 coordinates with a North 
Celestial Pole projection so the Parkes images were precessed, reprojected, 
and regridded to the same pixel size as the ATCA images.  The images from the 
two telescopes were then 
combined with the MIRIAD routine IMMERGE which Fourier transforms both images 
to obtain their spatial frequency components, compares them in the region of 
overlap, and then creates the merged image by transforming the combined data 
sets. 

After the precession, reprojection, and regridding, we also found that small 
translations of the Parkes images were needed to get the best alignment with 
the ATCA ones.  The 4.75 GHz image was shifted 8\arcsec ~east and 
3\arcsec ~north; the 8.55 GHz one was shifted 36\arcsec ~east and 
13\arcsec ~north.

To compare the integrated flux densities, we have summed them over the entire 
field of view shown in Figure 3 at 4.8 GHz but at 8.6 GHz the areas have been 
truncated to match the smaller area covered by the Parkes image at that 
frequency.  The integrated flux density on the Parkes image is 305.4 Jy and,
as expected, it is the same for the combined image.  The ATCA image alone has 
an integrated value of 32.4 Jy or about 11\% of the total.  This number is 
consistent with the conclusion of \citet{Haynes91} that about 40\% of the 
emission from the LMC comes from discrete sources because several of the 
large H~II region complexes, in particular 30 Doradus and N~11 in the 
northeast, cover areas of about 1/2$^{\circ}$, much larger than the response 
size of the compact Array. At 8.6 GHz, the Parkes value is 321.4 Jy and the 
merged image gives 325.7 Jy, acceptably close.  The integrated flux density 
from the ATCA is 18.8 Jy. or 6\% of the total.  We conclude that a significant 
fraction of the brightness of the LMC resides in features with scale sizes 
of order 10 arcmin, intermediate between the largest measurable sizes of the 
Compact Array at 8.6 GHz and 4.8 GHz.  

The merging process is not perfect.  Particularly at 8.6 GHz, there is some 
residual striping from the Parkes scanning pattern.  More importantly, 
particularly around a few bright sources 
with diameters of about 2\arcmin, it can be seen that they appear to lie in a 
ring with a small bowl in the center corresponding to the 30-meter spacing.  
This structure is most obvious toward N 132D, the brightest radio SNR in the 
LMC, at $05^h25^m05^s$ and $-69^{\circ}37'45''$.  Because of the intermediate 
size of these sources, there is insufficient power in the adjacent parts of 
the spatial frequency plane in the Parkes data to fully compensate for this 
pattern in the synthesized image. 

The normalization between the flux density scales for the Parkes observations 
in 1987-88 and the ATCA ones in 2001-03 was checked in two ways.  The first 
was to compare the adopted flux densities for the source PKS 1934$-$638 after 
corrections for the slight differences in frequency.  This comparison gave 
multiplying factors of 0.91 at 4.8 GHz and 1.09 at 8.6 GHz.  The second 
method was to use the procedure in IMMERGE that determines the average ratio 
of the visiblilites in the overlap region between the ATCA and the Parkes 
telescope which was chosen to be 25 -- 40 meters.  This operation gave 
values of 
0.88 at 4.8 GHz and 1.27 at 8.6 GHz.  At 4.8 GHz, the two values were very 
close and a multiplier of 0.9 was adopted.  To further evaluate the 
significant discrepancy between the values at 8.6 GHz, we attempted to 
minimize the bowl around N132D by trial and error.  This resulted in a 
multiplier of 1.10, essentially the same as the ratio of the official flux 
density scales and so was adopted.  We suggest that the average 8.6 GHz 
intensities are weak enough to be easily distorted by noise which produced a 
somewhat inaccurate ratio in the small region of overlapping visibilities 
sampled by IMMERGE.

\section{RESULTS}

\subsection{Images}

\begin{figure*}
\epsscale{1.0}
\plotone{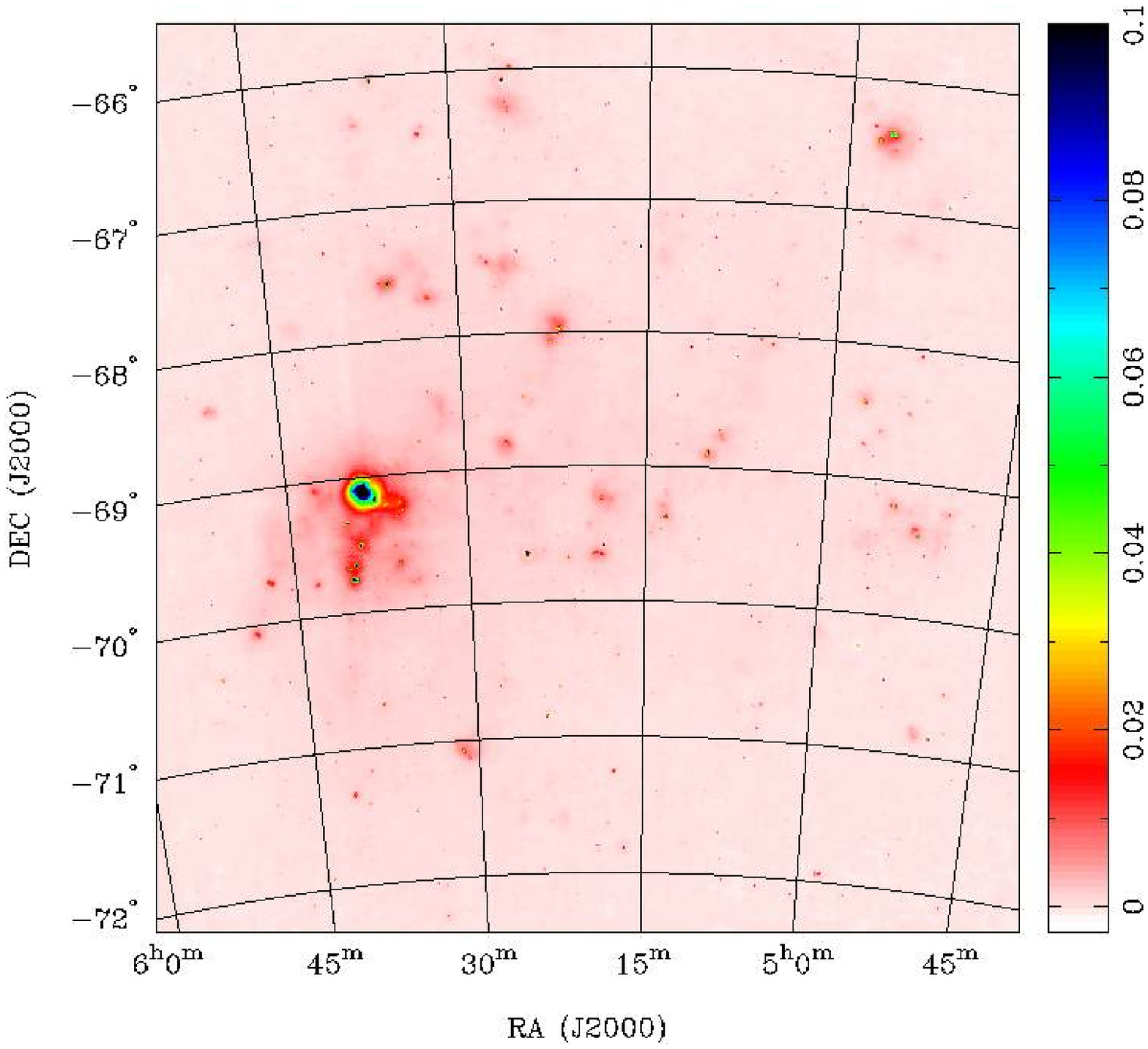}
\caption{Full image of the Large Magellanic Cloud at a frequency of 4.8 GHz 
with a half-power beamwidth of 33 arcsec.  The units on the wedge are Jy 
beam$^{-1}$.}
\label{fig-4800image}
\end{figure*}

\begin{figure*}
\epsscale{1.0}
\plotone{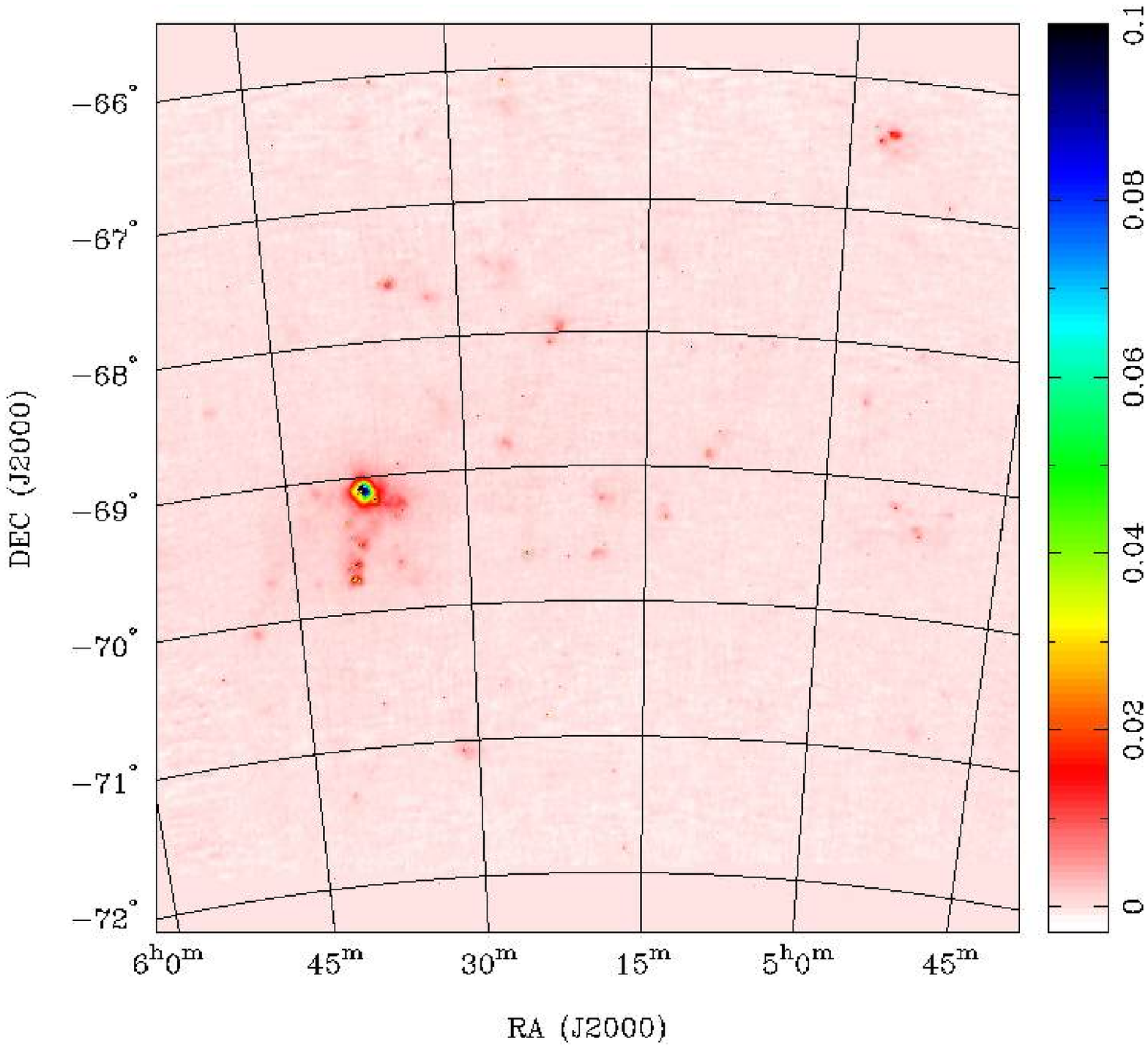}
\caption{Fullimage of the Large Magellanic Cloud at a frequency of 8.6 GHz 
with a half-power beamwidth of 20 arcsec.  The units on the wedge are Jy 
beam$^{-1}$.}
\label{fig-8640image}
\end{figure*}

The full merged images containing the data from the ATCA and Parkes are shown 
in Figures~\ref{fig-4800image} and \ref{fig-8640image}.
The full dynamic range at 4.8 GHz covers surface
brightnesses from about 0.0005 Jy 
beam$^{-1}$ to the peak of 30~Doradus at 1.3 Jy beam$^{-1}$ and so the color 
scale is a compromise to show as many interesting features as possible.  
Contours become too crowded to show details any better.  The smaller beamwidth 
at 8.6 GHz reduces the surface brightness for a uniform source to 0.37 times 
that at 4.8 GHz so the color scale has been adjusted accordingly. 
H~II regions should appear similar at both frequencies but supernova remnants 
and background sources are relatively fainter at 8.6 GHz.  Sometimes the 
ripples that show up in the 8.6-GHz image in multiples of the
$1\farcm 2$ scan increment make it difficult to see the relative brightnesses 
of the weak sources at the two wavelengths.
 
We have produced polarizarion images as well (see the data availability below) 
but do not show them here as virtually all the polarized emission bright 
enough to see is in fine scale features which are lost in the full-sized 
images.  As noted by \citet{Haynes91} even the Parkes data had to be smoothed 
to show detectable extended polarized structure.  The polarization data from 
Parkes are included in the images at 4.8 GHz but have little effect.   

In addition, to allow direct comparison, we have constructed images that have 
filtered the long spacings at 8.6 GHz and the short ones at 4.8 GHz  
to give the same range of spatial frequency response.  This match filtering in 
the $uv$-plane is particularly useful when determining spectral indices for 
extended sources.  At 8.6 GHz, this filtering is equibvalent to truncation of 
the $uv$ data at 8 k$\lambda$.  We shall hereafter refer to these images as 
``match filtered'' or ``match truncated.''  When merging with the Parkes data 
to get the full range of short spacings the filtering of the 4.8 GHz data is 
unnecessary.  In practice the difference between the filtered and 
simply convolved total intensity images at 8.6 GHz was negligible so for the 
polarization images we have simply used the convolved Stokes Q and U images.

\begin{figure}
\plotone{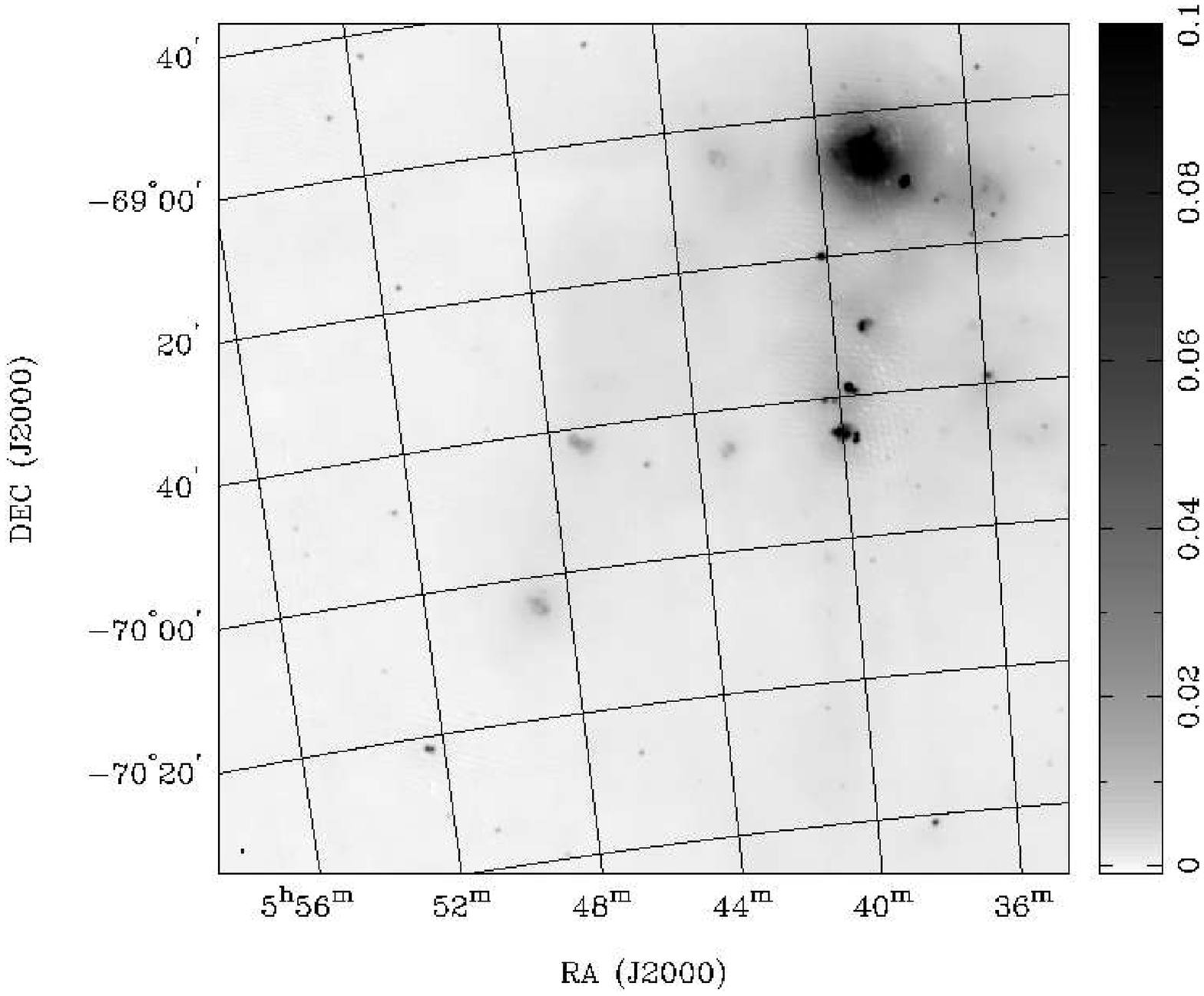}
\caption{A detailed figure showing the 4800 MHz emission from the 
region around 30~Doradus.  The units on the wedge are Jy beam$^{-1}$.  
The field of view is 1/16 of the
total LMC mosaic.  Figures~\ref{detail4800_fig}a-p show detailed 
subregions across the entire LMC but are presented in the electronic 
edition only.  Central coordinates for each subregion are given in
Table~1.}
\label{detail4800_fig}
\end{figure}

\begin{figure}
\plotone{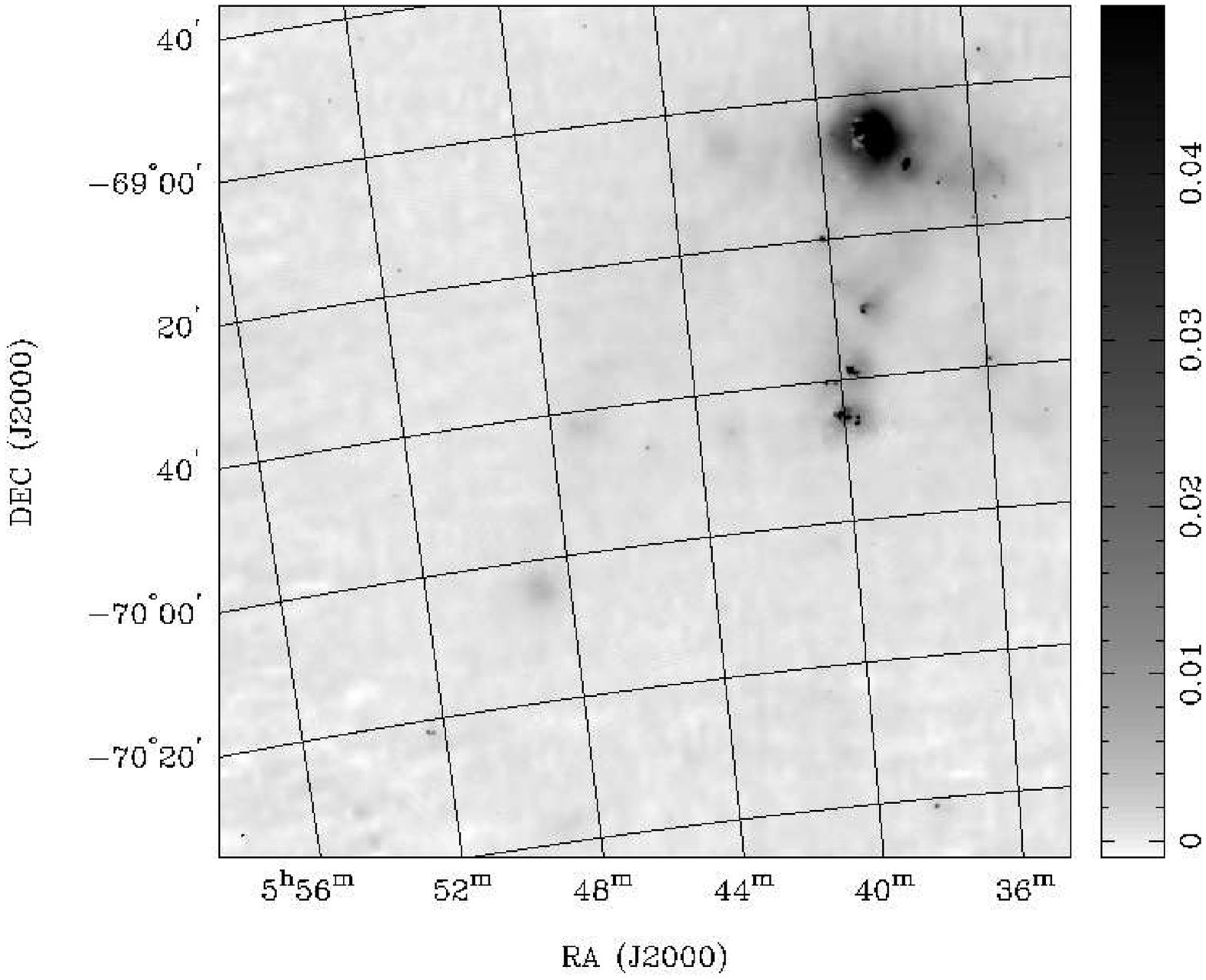}
\figurenum{6}
\caption{A detailed figure showing the 8640 MHz emission from the 
region around 30~Doradus.  The units on the wedge are Jy beam$^{-1}$.
The field of view is 1/16 of the
total LMC mosaic.  Figures~\ref{detail8640_fig}a-p show detailed subregions 
across the entire LMC but are presented in the electronic edition 
only.  Central coordinates for each subregion are given in Table~1.}
\label{detail8640_fig}
\end{figure}

Subimages are, of course, needed to see details of any area.  We show sample 
images at 4.8- and 8.6 GHz of the area including the giant HII 
region - molecular cloud complex 
30~Doradus in Figs.~\ref{detail4800_fig} and \ref{detail8640_fig},
respectively.   The electronic  
version of the figures includes 16 total-intensity subimages at each frequency 
with 1450 
pixels on a side and a 300 pixel overlap for each image. The central
coordinates of these sub-images are listed in Table~1. They all have the 
same NCP projection and reference position at $05^{h}19^{m}00^{s}$ and 
$-69^{\circ}00'00''$.  

\begin{figure*}
\epsscale{0.95}
\plotone{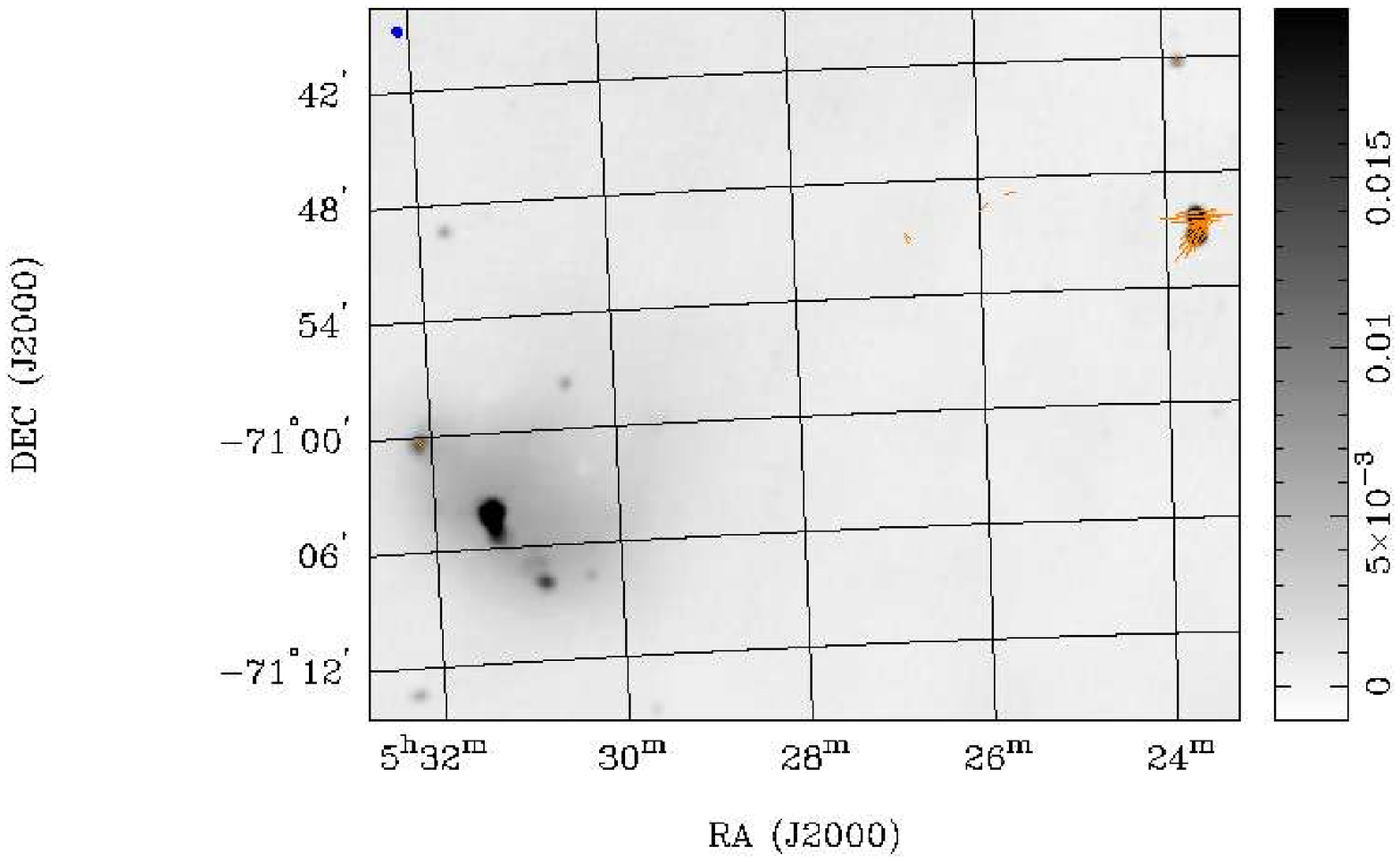}
\plotone{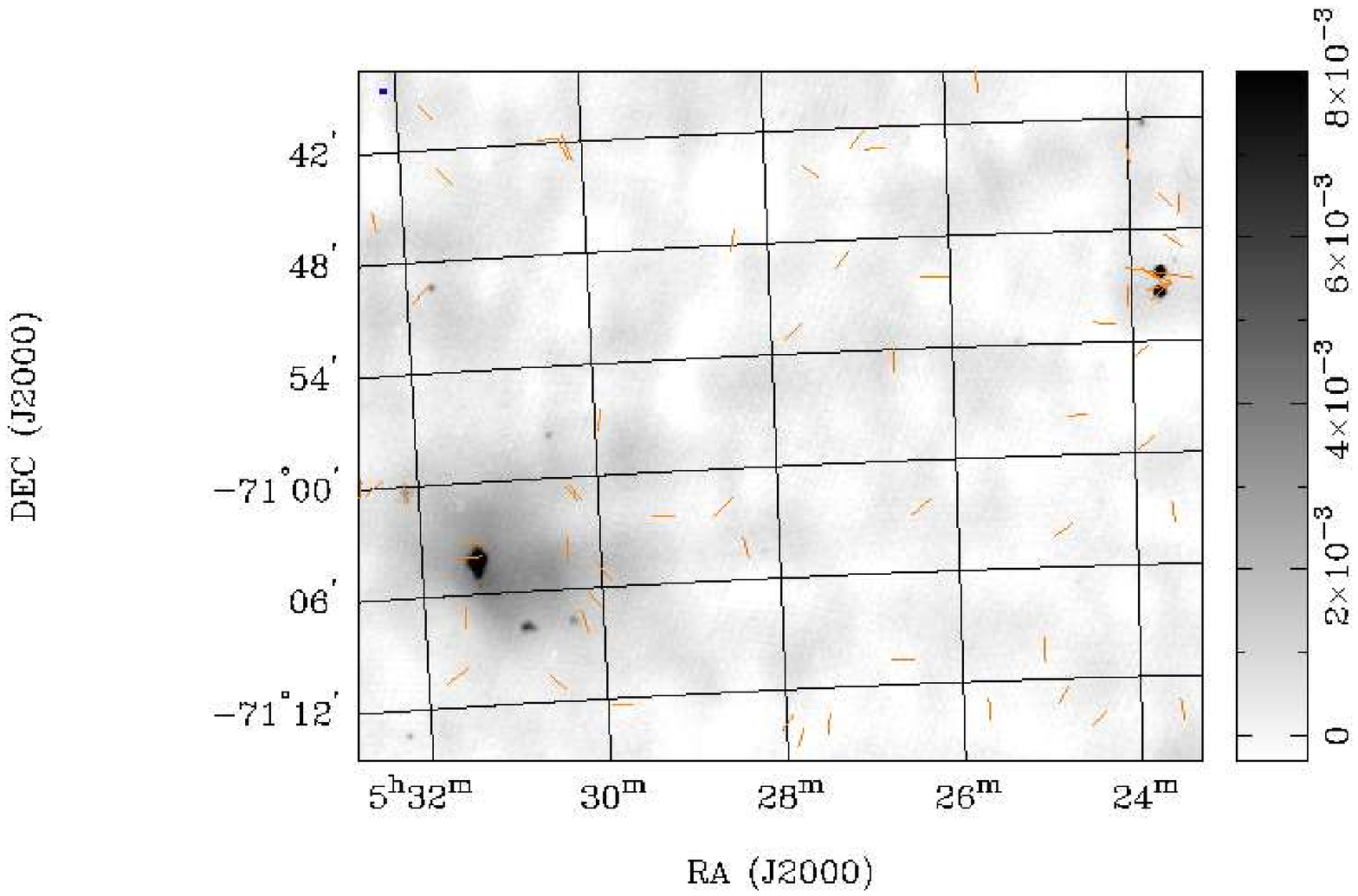}
\caption{Images of the region containing the H~II region 
and SNR complex N206 plus two small diameter sources.  The greyscale is the 
total intensity with the units on the wedge in Jy beam$^{-1}$.  The lines are 
the polarized electric vectors.  The beam is 
shown in the upper left corner.  a) 4.8 GHz where the longest vector 
represents a polarized intensity of 8 mJy beam$^{-1}$. b) 8.6 GHz where the 
longest vector is 4 mJy beam$^{-1}$ .}
\label{fig-region1}
\end{figure*}

To illustrate the details that can be determined from more specific subimages 
showing every 
pixel, we show results on two sample areas.  Figures~\ref{fig-region1}~a,b are 
total intensity images at both frequencies with polarization e-vectors 
superposed of an area containing the H~II 
region/SNR complex N206 on the east side at about $05^{h}31^{m}$ and 
$-71^{\circ}05'$  and an apparent double source on the west side at 
$05^h23^m41^s$ and $-70^{\circ}51'23''$. These and all subsequent images shown,
both total intensity and polarization, are merged ATCA plus Parkes images.  
The polarized intensities are all truncated at 4 $\sigma$ so that most of the 
remaining vectors are real.  The SNR is on the 
northeast side of N206 and is more prominant in the 4.8-GHz image than in 
the 8.6-GHz one because of its non-thermal spectrum.  Although the 
polarimetric data are somewhat noisy at 8.6 GHz, we can see that the brightest 
part of the SNR is polarized  whereas the larger H~II region has only random 
vectors at about the 3-$\sigma$ level on the 8.6 GHz image.  The SNR has 
already been investigated in more detail by \citet{k02}.  

The two small 
diameter objects on the western edge of Fig.~\ref{detail8640_fig} can
just be discerned in the  
full images in Figures~\ref{fig-4800image} and \ref{fig-8640image}.
They have been reported as a 
single source in all previous catalogs (e.g. Filipovic et al. 1995).  They 
are both significantly polarized at each frequency and 
assuming no 180$^{\circ}$ ambiguity (1163 rad m$^{-2}$) between the two 
frequencies, we determine mean Faraday rotations of 
+71 rad m$^{-2}$ for the northern source and $-$26
rad m$^{-2}$ for the southern one.  These small values are typical of 
those found for the LMC in general \citep{k93}.  

\begin{figure*}
\epsscale{1.1}
\includegraphics[angle=270,scale=.7]
{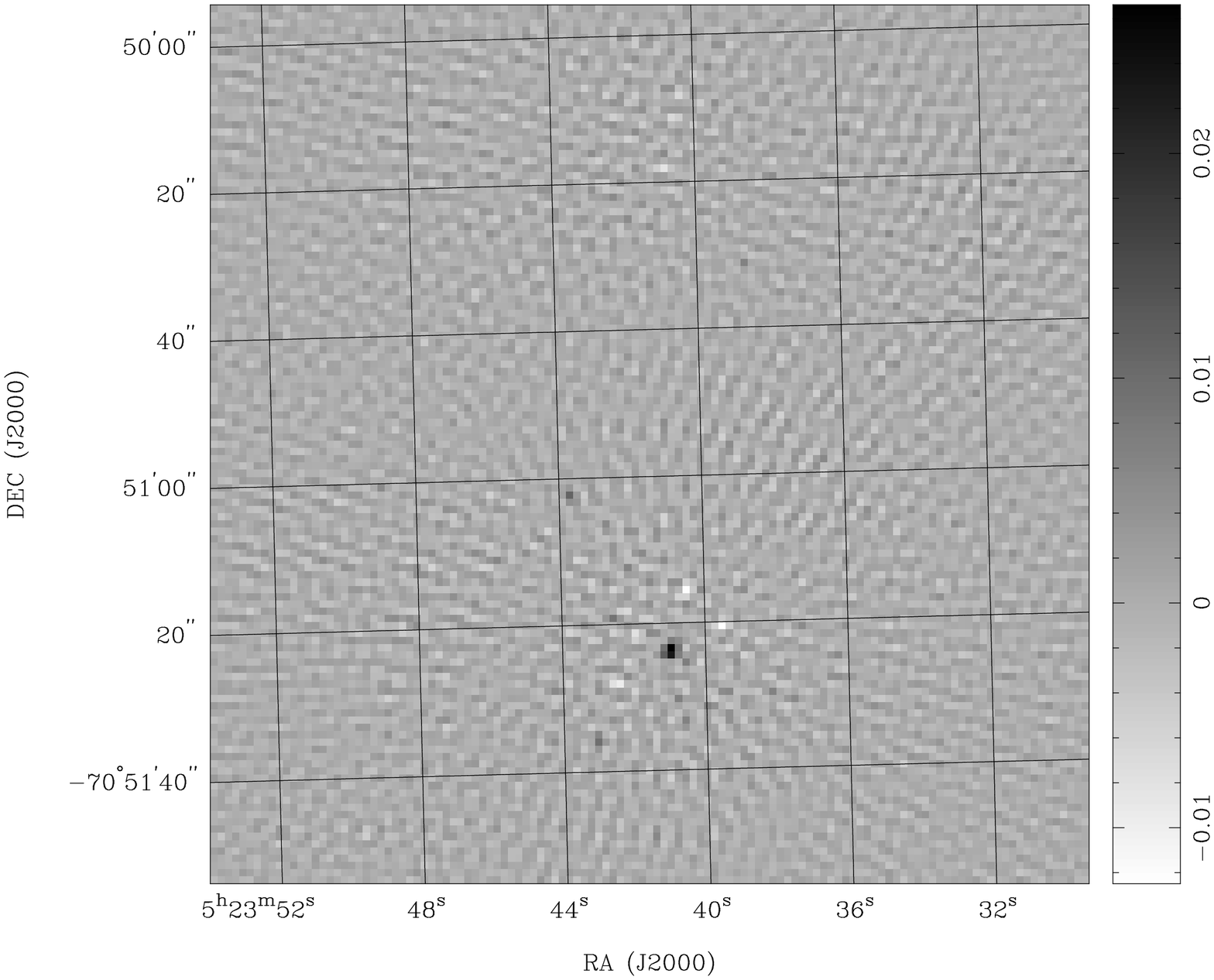}
\caption{High resolution image at 4.8 GHz of the two small diameter sources in 
Figure~\ref{fig-region1}.  It is constructed from data taken only with
the 6-km antenna 
correlated with each of the others so that it is sensitive only to features 
with a size of $<$ 3 arcsec.  The units on the wedge are Jy 
(3-arcsec beam)$^{-1}$.  This shows that the southern source is 
point-like while the missing northern one that should be at 
$05^{h}23^{m}41^{s}$ and $-70^{\circ}50'18''$ is resolved out showing that it 
is extended.}
\label{fig-region1.6km}
\end{figure*}

Neither source significantly broadens 
the 20\arcsec ~beam at 8.6 GHz, and because of the weak source intensity in 
the presence of the noise, it is not possible to evaluate the angular size of 
either object.  To gain some information on a scale of a few arcsec, we show 
a  4.8-GHz image in Fig.~\ref{fig-region1.6km} made with data
only from antenna 6 correlated with each of the others so it shows only the 
emission on scales smaller than about 3 arcsec.  The southern source remains 
unresolved and is likely an extragalactic background source.  The northern 
source is obviously extended at that resolution, probably with a size of 
5 - 10 arcsec, and could be a new 
small-diameter SNR in the LMC.  Unfortunately, the ROSAT All Sky Survey 
has a gap around that region so we have not yet been able to follow up 
on that source.  But, in that quick look we have been able to 
identify a likely background source for which we can measure the Faraday 
rotation through the LMC and have possibly identified a new SNR. In addition 
to all the definitely extended LMC sources in the images, there are about 
three dozen other small sources for which similar studies can be done. 

\begin{figure*}
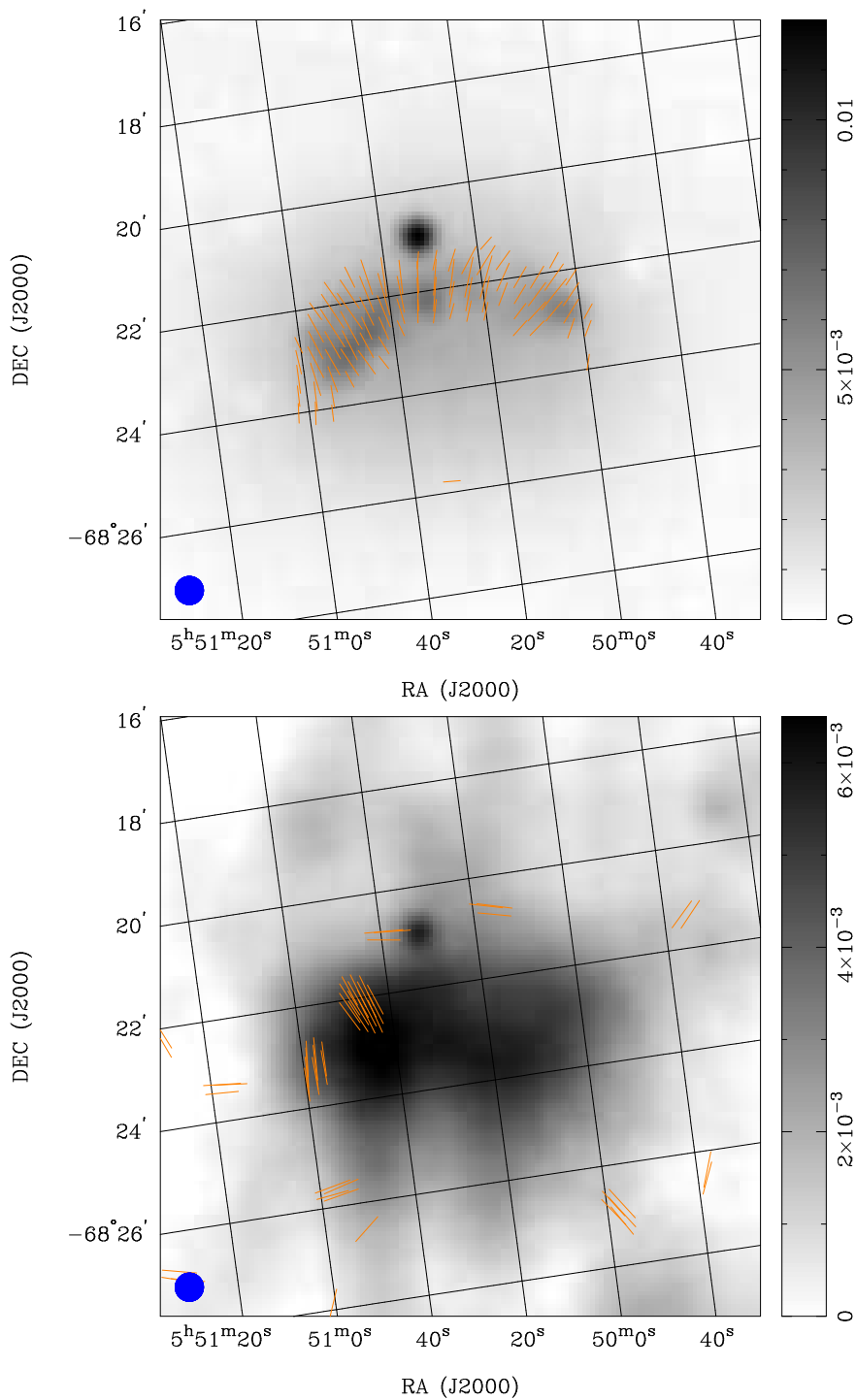

\epsscale{0.7}
\plotone{dickel.fig9a.cps}
\plotone{dickel.fig9b.cps}
\caption{Images of the area around a supernova remnant and extragalactic 
background source in the LMC at a)  4.8 GHz and b) 8.6 GHz. The greyscales are 
the total intensities with the units on the wedges in Jy beam$^{-1}$.  
The lines represent the polarization electric 
vectors.  The 8.6-GHz image is shown with the 
same resolution as the 4.8-GHz one for direct comparison. The half-power 
beamwidth is shown in the bottom left corners.}
\label{fig-0550}
\end{figure*}

\begin{figure*}
\epsscale{1.1}
\plottwo{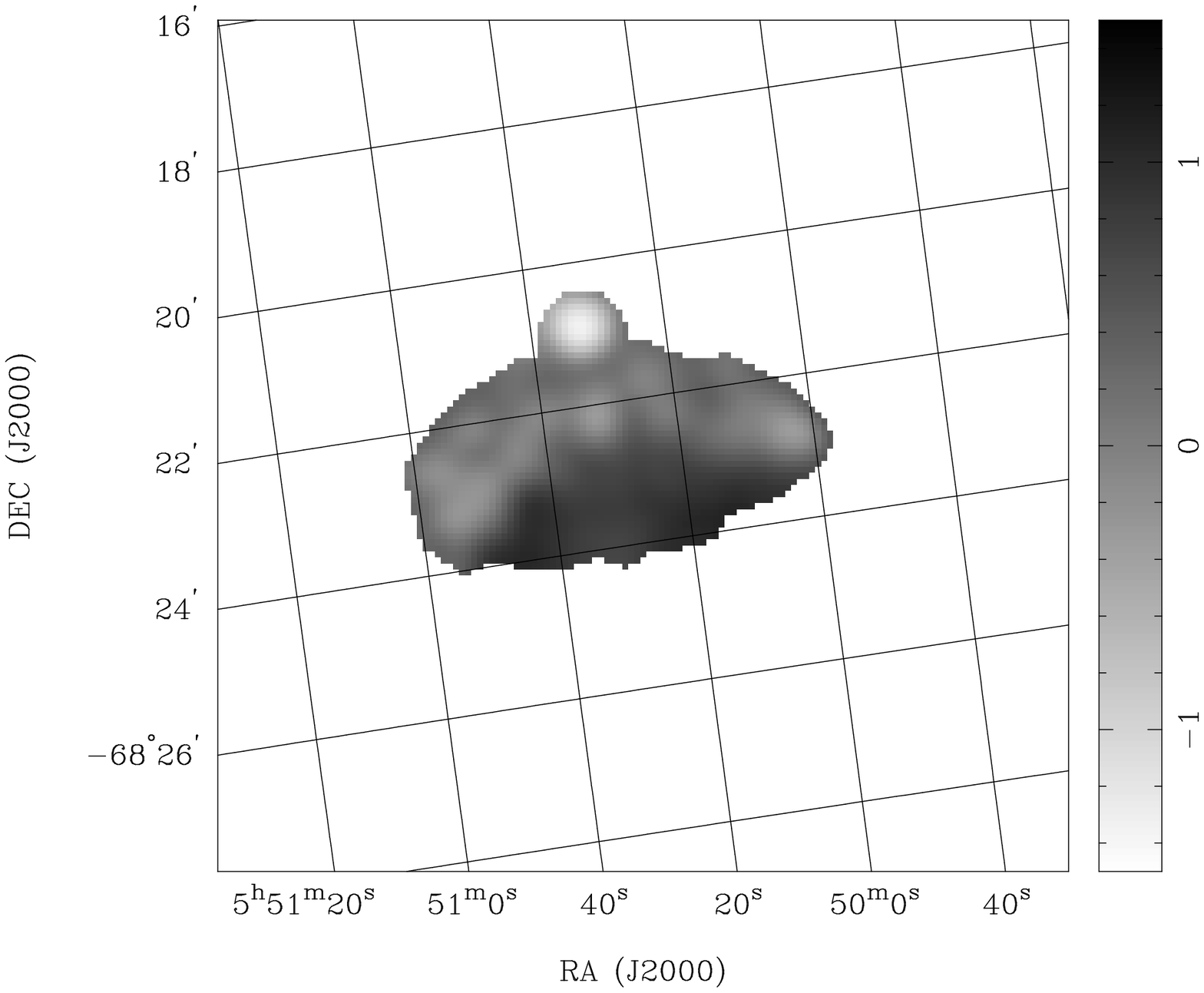}{dickel.fig10b.cps}
\caption{a)  Spectral index map of a supernova remnant and background source 
in the LMC.  The units on the wedge are the spectral index, $\alpha$, as 
defined in the text.  b)  Total intensity map at 4.8 GHz with superposed 
vectors representing the magnetic field directions and the polarized intensity 
at 4.8 GHz.  The units on the wedge are Jy beam$^{-1}$.}
\label{fig-0550spix} 
\end{figure*}

A second interesting area is on the eastern edge of the LMC at $05^{h}50.5^{m}$
and $-68^{\circ}22'$.  In Figure~\ref{fig-0550}~a,b we see a polarized
curved arc which  
trails off to the south and a small-diameter source just to the north of the 
center of the arc.  The north source slightly broadens the 20\arcsec ~response 
of the beam at 8.6 GHz suggesting that it has a diameter of 10 $\pm$ 5 
arcsec. 
Figure~\ref{fig-0550spix} shows the result of determining the spectral
index from the match-filtered
images at 4.8 and 8.6 GHz.  Where the arc is bright enough to be 
above the noise on the southern edge, it has a spectral 
index, $\alpha$, of about $-$0.4 
(where the flux density $S_f \propto f^{\alpha}$) whereas the north source has 
a value near $-$1.0.  The former is typical of old shell supernova remnants
($http://cats.sao.ru/snr\_spectra.html$)
whereas the latter is characteristic of extragalactic sources.  We note that a 
spectral index determination between the lower-resolution MOST image at 843 
MHz and a smoothed 4.8-GHz image give the same spectral index results.  

Using the 
polarimetry to determine the Faraday rotation and intrinsic direction of the 
magnetic fields, we find that the mean Faraday rotation changes rather rapidly 
from $-$100 rad m$^{-2}$ to +150 rad m$^{-2}$ on the eastern edge of the SNR 
shell but then remains relatively constant at about +50 rad m$^{-2}$ across 
the rest of the arc. The magnetic field directions are shown in
Figure~\ref{fig-region1.6km}. It 
appears that the magnetic field is approximately radial on the eastern edge 
and then tangential around the the rest of the arc.  

The X-ray emission from this 
region observed by Chandra \citep{w04} matches the radio emission in the arc 
quite closely and shows a point at the position of the small-diameter north 
source.  A long filament, 
DEM L238, seen faintly in H$\alpha$ extends from about 1$^{\circ}$ east of 
this through the area between the arc and the north source and trailing off 
about 10 arcmin further west.  Somewhat brighter H$\alpha$ and [S~II] emission 
are seen to the south of the arc.  We conclude that 
the north source is probably an extragalactic background source and that the 
arc is a supernova remnant encountering extra material on its northern side 
where the shock speed is slower.  
 
\subsection{Data Availability}

A variety of images and the $uv$ data are available in FITS format from the 
Astronomical Digital Imaging Library (ADIL) at the National Center for 
Supercomputing Research at the University of Illinois 
(http://adil.ncsa.uiuc.edu/document/04.JD.01).  The available data are listed 
in Table 2. 

\subsubsection{Images}

For consistency, all of the images are of the full galaxy with 4624 $\times$ 
4872  5\arcsec ~pixels.  The displayed projection will cause some blank pixels 
around the edges and the valid image at 8.6 GHz will be slightly smaller than 
that at 4.8 GHz because of the smaller primary beam at the higher frequency.  

Because most users will be interested in particular areas where more detail 
can be seen, subimages will also be available in the near future.  The user 
will be able to request a central 
position and the size of the image.  They will retain the 5\arcsec ~pixels but 
a reference position of the center of the subimage.  

The images include the full resolution total intensity at both 
frequencies for the ATCA data alone and for the merged ATCA and Parkes data 
sets. We also include the match filtered total intensity image 
both merged and unmerged at 8.6 GHz. 

For polarization, Stokes Q and U images are available at both frequencies 
along with polarized intensity and position angle ones. The latter have been 
corrected for bias but not cutoff at any level so they are appropriate for 
fractional polarization determinations but any individual points below 3-4 
times the rms noise level cannot be trusted.  Both full resolution and 
convolved polarimetric data are available at 8.6 GHz.

Polarization data from Parkes were available only at 4.8 GHz and merged 
polarimetric images were made at that frequency.  In the 
direction of 30~Dor, the polarizarion leakage at Parkes can reach about 
0.7\% so the merged polarimetry cannot be used for that source but should 
be ok for the rest of the galaxy.

\subsubsection{Visibility Data}

The calibrated $uv$ data are also available from ADIL in FITS format.  They will
contain all polarizations and include the long-spacing data from antenna 6.  
The entire data set may be downloaded in a single separate FITS file for each 
of the two frequencies. Each file will contain 
thirteen 8-MHz channels centered on the nominal frequency.

\section{FUTURE USES OF THE DATA}

Some of the many things that can be done with these data are listed below.  
We and co-workers are doing some of them but readers are encouraged to obtain 
the images from ADIL to use them as they wish.
    
\begin{itemize}
\item Measure flux densities and spectral indices of extended sources and
classify them, making use of optical, X-ray and existing radio datasets.
Reliable continuum spectral indices require an extended
frequency baseline with adequately matched $uv$ coverage.
By comparing to the MOST data as well (45$\arcsec$ HPBW at 843 
MHz) as well \citep{t98,mg99} we cover a decade in frequency, which
should be enough to distinguish SNRs from H~II regions and provide
some discrimination among non-thermal spectral indices.

\item Catalogs of the sources will be used to evaluate statistical 
luminosities, energies, birthrates, etc. of massive stars and their SNe in
the LMC.

\item The typical size of known SNRs in the LMC is 
30\arcsec -- 8\arcmin ~\citep{w99}. Previous radio surveys do not have adequate
resolution
for comparison of the X-ray and radio morphology, but the proposed
observations
will. The data will enable a study of the bivariate luminosity function of
SNRs.

\item H~II regions may follow a power-law size distribution
\citep{kh86} with most smaller than 100 pc (almost 7 arcmin) in diameter.
The proposed observations will enable detection of internal structures 
in most of these to 
evaluate their excitation.  With the exception of 30~Doradus, all the  
H~II regions in the catalogs of \citet{h56} and of \citet*{Davies76} have been 
resolved for the first time in the radio.
   
\item The ROSAT X-ray mosaics of the Magellanic Clouds have revealed
diffuse X-ray emission in regions with scale sizes of 10 -- 10$^3$
pc  \citep{s99b}.  The larger sources include
superbubbles (10$^2$ pc), supergiant shells (10$^3$ pc), and
unconfined fields ($\sim$10$^3$ pc).  The proposed continuum
survey will allow us to search for nonthermal radiation
associated with the diffuse X-ray emission regions and study
the generation of relativistic electrons by supernova remnant
shocks in a variety of environments including molecular clouds and 
H~I shells.

\item The full polarimetric capability can be used to obtain valuable 
Faraday rotation information on the LMC and sources within it.  In 
addition to the SNRs such as DEM L 316, DEM L 328, and N 132D, about three 
dozen unresolved sources have measurable polarization.  No mapping of the 
Faraday rotation of background sources through the LMC has been previously 
done.

\item Data from baselines including the 6~km antenna will be used to extend the
catalog of unresolved objects \citep*{m97} in terms of
spatial coverage, angular resolution, frequency range and sensitivity.
The proposed survey will provide more accurate positions for
extragalactic radio sources (e.g., AGNs and QSOs), and produce
a list of UV- and optical-bright objects, which can be used
as probes for interstellar absorption line observations of
physical conditions and abundances of the LMC's ISM.

\end{itemize}

The Australia Telescope Compact Array is part of the Australia
Telescope funded by the Commonwealth of Australia 
for operation as a National Facility, managed by CSIRO.  We have benefitted 
from help and valuable discussions with many colleagues.  They include Bob 
Sault, Lister Staveley-Smith, John Dickey, Uli Klein, H{\'e}l{\`e}ne Dickel, 
Robin Wark, Ray Plante, You-Hua Chu, Rosa Williams, and Brian Fields.  Barry 
Parsons, Margaret House, and 
Vicki Drazenovic helped to make the many visits to the Australia Telescope 
very enjoyable.  We appreciate valuable comments by the referee, Richard 
Wielebinski.  JRD acknowledges support from the Campus Honors Program of 
the UIUC and NASA Grant NAG5-11159.  RAG acknowledges support from NSF Grant 
AST-0228953 to BIMA.



\begin{deluxetable}{cllcc}
\tablewidth{0pt}
\tablecaption{Field Centers for Subimages}
\tablehead{
\colhead{Fig.} & \colhead{Field} & \multicolumn{2}{c}{Center Location} \\
\colhead{Id\tablenotemark{a}}  & \colhead{Name}  & \colhead{$\alpha$(J2000)} & 
\colhead{$\delta$(J2000)}
}
\startdata
a & nnee  &  5$^{\rm h}$41$^{\rm m}$54\fs 30  & $-$66\degr 33\arcmin 58\farcs 92
 \\
b & nne   &  5$^{\rm h}$27$^{\rm m}$07\fs 59  & $-$66\degr 40\arcmin 54\farcs 08
 \\
c & nnw   &  5$^{\rm h}$12$^{\rm m}$16\fs 31  & $-$66\degr 41\arcmin 36\farcs 81
 \\
d & nnww  &  4$^{\rm h}$57$^{\rm m}$27\fs 90  & $-$66\degr 36\arcmin 06\farcs 67
 \\
e & nee   &  5$^{\rm h}$43$^{\rm m}$33\fs 67  & $-$68\degr 09\arcmin 36\farcs 20
 \\
f & ne    &  5$^{\rm h}$27$^{\rm m}$45\fs 79  & $-$68\degr 16\arcmin 55\farcs 04
 \\
g & nw    &  5$^{\rm h}$11$^{\rm m}$52\fs 32  & $-$68\degr 17\arcmin 40\farcs 23
 \\
h & nww   &  4$^{\rm h}$56$^{\rm m}$02\fs 36  & $-$68\degr 11\arcmin 51\farcs 21
 \\
i & see   &  5$^{\rm h}$45$^{\rm m}$27\fs 77  & $-$69\degr 44\arcmin 05\farcs 30
 \\
j & se    &  5$^{\rm h}$28$^{\rm m}$29\fs 72  & $-$69\degr 51\arcmin 51\farcs 87
 \\
k & sw    &  5$^{\rm h}$11$^{\rm m}$24\fs 72  & $-$69\degr 52\arcmin 39\farcs 93
 \\
l & sww   &  4$^{\rm h}$54$^{\rm m}$24\fs 09  & $-$69\degr 46\arcmin 28\farcs 79
 \\
m & ssee  &  5$^{\rm h}$47$^{\rm m}$39\fs 98  & $-$71\degr 17\arcmin 26\farcs 51
 \\
n & sse   &  5$^{\rm h}$29$^{\rm m}$20\fs 71  & $-$71\degr 25\arcmin 45\farcs 76
 \\
o & ssw   &  5$^{\rm h}$10$^{\rm m}$52\fs 68  & $-$71\degr 26\arcmin 37\farcs 22
 \\
p & ssww  &  4$^{\rm h}$52$^{\rm m}$30\fs 16  & $-$71\degr 20\arcmin 00\farcs 00
 \\
\enddata

\tablenotetext{a}{The letter refers to the subfigure identification for each field
in Figures 5 and 6 of the electronic edition.}

\end{deluxetable}

\begin{deluxetable}{lccc}
\tablewidth{0pt}
\tablecaption{FITS Data Available}
\tablehead{
\colhead{Name}  & \colhead{Freq. (MHz)} & \colhead{Type} & \colhead{Other Attributes} \\
\\
\colhead{Images}}
\startdata
LMC4.8-i.a & 4.8 & total intensity & only ATCA \\
LMC4.8-i.m & 4.8 & total intensity & merged ATCA + Parkes \\
LMC4.8-i.f & 4.8 & total intensity & match filtered, only ATCA \\
LMC4.8-q.a & 4.8 & Stokes Q & only ATCA \\
LMC4.8-q.m & 4.8 & Stokes Q & merged ATCA + Parkes \\
LMC4.8-u.a & 4.8 & Stokes U & only ATCA \\
LMC4.8-u.m & 4.8 & Stokes U & merged ATCA + Parkes \\
LMC4.8-p.a & 4.8 & polarized intensity & only ATCA \\
LMC4.8-p.m & 4.8 & polarized intensity & merged ATCA + Parkes \\
LMC4.8-pa.a & 4.8 & e-vector position angle & only ATCA \\
LMC4.8-pa.m & 4.8 & e-vector postion angle & merged ATCA + Parkes \\ 
\\
LMC8.6-i.a & 8.6 & total intensity & only ATCA \\
LMC8.6-i.m & 8.6 & total intensity & merged ATCA + Parkes \\
LMC8.6-i.f & 8.6 & total intensity & match filtered, only ATCA \\
LMC8.6-i.f.m & 8.6 & total intensity & match filtered, merged \\
LMC8.6-q.a & 8.6 & Stokes Q & only ATCA \\
LMC8.6-q.c & 8.6 & Stokes Q & convolved \\
LMC8.6-u.a & 8.6 & Stokes U & only ATCA \\
LMC8.6-u.c & 8.6 & Stokes U & convolved \\
LMC8.6-p.a & 8.6 & polarized intensity & only ATCA \\
LMC8.6-p.c & 8.6 & polarized intensity & convolved \\
LMC8.6-pa.a & 8.6 & e-vector position angle & only ATCA \\
LMC8.6-pa.c & 8.6 & e-vector position angle & convolved \\
\hline
\\
~~~UV data & & &  \\
\hline
LMC4.8-uv & 4.8 & full Stokes & calibrated, all 6 antennas  \\
LMC8.6-uv & 8.6 & full Stokes & calibrated, all 6 antennas \\
\enddata
\end{deluxetable}


\end{document}